\documentclass{kapproc} 

\usepackage{t1enc}

\usepackage{procps} 

\usepackage[dvips]{graphicx}

\upperandlowercase

\kluwerbib 

\begin{document}

\articletitle{Evolution of Thermal Conducting Clouds Embedded in
              Galactic Winds}

\author{A. Marcolini,\altaffilmark{1} D. K. Strickland,\altaffilmark{2}
        A. D'Ercole\altaffilmark{3} and T. M. Heckman \altaffilmark{2}}

\altaffiltext{1}{Dipartimento di Astronomia, Universit\`a di Bologna,
       via Ranzani 1, 44127 Bologna, Italy}
\email{andrea.marcolini@bo.astro.it}

\altaffiltext{2}{Physics and Astronomy Department, Jhons Hopkins University,
            Homewood Campus, \\
            Baltimore, MD 21218, USA}

\altaffiltext{3}{Osservatorio Astronomico di Bologna,
       via Ranzani 1, 44127 Bologna, Italy}

\begin{abstract}
We performed high resolution hydrodynamical simulations of dense cool
clouds embedded in supernova-driven galactic superwinds. Here we
present preliminary results of our reference model in which the
effect of heat conduction are taken into account.
Significant dynamical differences occur between simulations with and
without heat conduction. In absence of heat conduction the cloud
fragments in few dynamical timescale. The inclusion of heat
conduction has the effect to stabilize the cloud and inhibit the
growth of Kelvin-Helmholtz and Reyleigh-Taylor
instabilities. Furthermore in the conditions met in our simulations the
strong heat flux at the cloud edge generates a converging
shock which compresses the cloud.
We also calculate the high energy emission (O{\sc vi} and soft
X-ray) of the cloud and O{\sc vi} absorption line properties and compare the
results with obvervations.  Models in which heat conduction is taken
into account seem to fit the observations much better. In general only
a small fraction (0.1-0.4\%) of the wind mechanical energy intersecting 
the cloud is radiated away.
Finally some of our models are able to explain the low metallicity
abundance seen in X-ray osservation of superwinds.

\end{abstract}

\section{Introduction}

Superwinds are multi-phase, loosely-collimated galaxy-sized outflows
with measured velocities in excess of several hundred to a thousand
kilometers per second, driven from galaxies experiencing intense
recent or ongoing star-formation, i.e. starburst galaxies. Superwinds
are believed to be driven by the thermal and ram pressure of an
initially very hot ($T\sim10^8$ K), high pressure ($P/k \sim 10^7$ K
cm$^{-3}$) and low density wind created from the merged
remnants of very large numbers of core-collapse supernovae (SNe). The
thermalized SN and stellar wind ejecta predicted by this model is too
hot and too tenuous to be easily observed. However hydrodynamical models
of superwinds show that the wind fluid sweeps up and incorporates
larger masses of ambient galactic disk and halo interstellar medium
(ISM) into the superwind, material which is more easily detected
observationally. Recent observations of O{\sc vi} absorption and
emission from $T\sim 10^{5.5}$ K gas in the far Ultraviolet, and
thermal emission from $T\sim 10^{6-7}$ K gas in X-ray regime, have
shown that the majority of the soft X-ray emission in superwinds is
not due to the wind fluid itself, but arises from some form of
interaction between the wind gas and denser ambient disk or halo
ISM. Here we present preliminary results of
high-resolution hydrodynamical simulations of the interaction of a
superwind with an embedded cool cloud. In this model we include the
effect of thermal conduction, which we show plays an important role in
shaping both the dynamics and radiative properties of the resulting
wind/cloud interaction. We compare the simulated O{\sc vi} and soft
X-ray properties to the existing observational data.

\section{Physical Assumptions and Computational Method}

In our models we make the following simplifying assumptions: the
clouds are initially spherical, at rest and in pressure equilibrium
with the ambient gas; magnetic fields and cloud self-gravity are
neglected. \par We run several models with different values of the
wind temperature $T_{\rm w} (10^6-10^7$ K), wind hydrogen number
density $n_{\rm H,w}=n_{\rm w}= (4.1\times 10^{-3} - 8.2\times
10^{-4}$ cm$^{-3})$ and wind velocity $v_{\rm w}=(447-2236)$ km
s$^{-1})$. Here we present results relative to the model with $T_{\rm
w}=5\times 10^6$ K, $n_{\rm w}=8.2\times 10^{-4}$ cm$^{-3}$, $v_{\rm
w}=1000$ km s$^{-1}$. The cloud properties are the same in all models:
radius $R_{\rm c}=15$ pc, temperature $T_{\rm c}=10^4$ K and density
$n_{\rm H,c}=n_{\rm c}=0.41$ cm$^{-3}$ (total mass $\sim 206
M_{\odot}$).\par We integrated the hydrodynamical equations with a 2D
hydrocode developed by the Bologna group. It is based on an explicit,
second order upwind method which makes use of the van Leer
interpolation method and of the consistent advection. The method
operates on a staggered grid and is implemented with thermal
conduction. Following Cowie and McKee (1977) we adopt saturated fluxes
to avoid unphysical heat transport in presence of steep temperature
gradients. The model presented here is run on a grid with $1200\times
400$ mesh points, with a central uniform region of $1000\times 300$
points and a mesh size $\Delta R=\Delta z=0.1$ pc, while in the outer
regions the mesh size increases logarithmically.\par Calculations of
the O{\sc vi} and soft X-ray luminosities and 2-dimensional maps of
the volume emissivities from these models were performed separately
from the simulations themselves. We assumed that the plasma is in
collisional ionization equilibrium. The O{\sc vi} emissivities we use
are based on the MEKAL hot plasma code (e.g. Mewe et al 1985). Note
that the luminosity quoted is the sum of the two lines in the
$\lambda=1032$ \AA~ and 1038 \AA~ doublet. The soft X-ray luminosities
we quote are in the 0.3-2.0 keV energy band, chosen to correspond to
the energy band used in the $Chandra$ ACIS observations we compare
to. The X-ray emissivities used are based on the 1993 update to the
Raymond \& Smith (1977) hot plasma code. For convenience we calculated
all luminosities and volume emissivities assuming Solar abundances (as
given in Anders \& Grevesse 1989), which is the commonly used standard
in X-ray astronomy), although the cloud and wind material may well
have different metal abundances.\par
\medskip

\section{Hydrodynamics}

As the wind moves supersonically (Mach number = 3.2) a bow shock forms
around the cloud, while a transmitted shock is driven into the
cloud. In absence of heat conduction such an interaction leads to a
fragmentation of the cloud due to Kelvin-Helmholtz (K-H) and
Rayleigh-Taylor (R-T) instabilities (e.g. Klein et al. 1994; Fragile
et al 2004) in a typical timescale $\sim (n_{\rm c}/n_{\rm w})^{1/2}
R_{\rm c}/v_{\rm w}$ ($\sim 3.3 \times 10^5$ yr for this model). 

The inclusion of heat conduction has the effect of stabilizing the
cloud, as it has the tendency to smooth out the temperature and density
gradients such that the K-H instability is noticeably reduced. The
heat conduction flux is also responsible for the reduction of the R-T
instability.

\begin{figure}[ht]
\includegraphics[width=\textwidth]{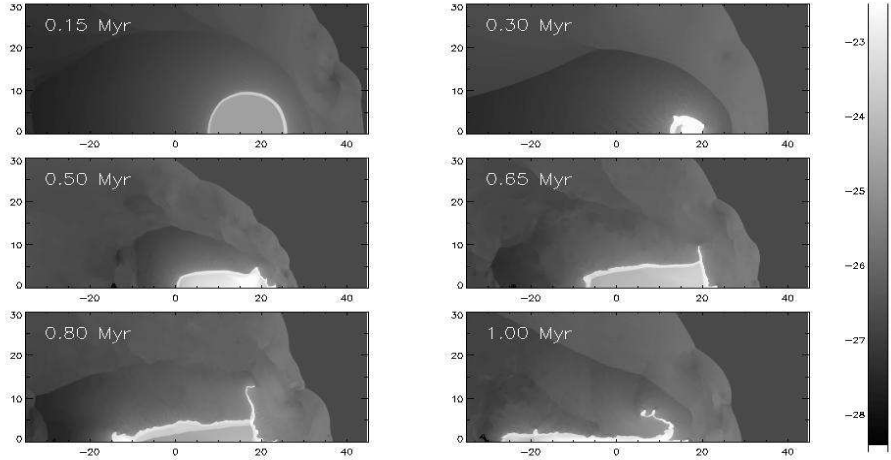}
\vskip.1in
\caption{Density distribution at different times of the cold cloud
interacting with the hot tenuous superwind including the effects of
heat conduction. The logarithm of the mass density (in units of
g cm$^{-3}$) is shown.  The distances are given in pc. At the beginning of
the simulation the cloud centre is at $z=20$ pc.}
\end{figure}
In fact, in the conditions met in our simulations, the strong heat
flux at the cloud edge generates a converging shock which compresses
the cloud (Cowie and McKee 1977); the reduction of the cloud size
implies a reduction of the momentum transfer from the superwind to the
cloud, and thus of the cloud acceleration responsible of the R-T
instabilities. As a consequence the cloud remains more compact with
bound edges.\par The behavior of the cloud is shown in Fig. 1. After
0.15 Myr the cloud is still roughly spherical; however the formation
of the converging shock at its edge is clearly visible. At 0.30 Myr
the cloud reaches its minimum size and then re-expands mostly
downstream where it is not contrasted by the effect of the ram
pressure of the incoming superwind. 
The shrinking of the cloud leads to a reduction the evaporating 
rate $\dot M$ that varies in pace with the radius.
After 1 Myr the cloud retains half of its initial mass while the 
rest has evaporated and mixed with the
superwind material.

In general we found that due to dynamical effect, the lifetime of 
our evaporating clouds is 2-3 times longer than that 
provided by the analytical solution of a steady evaporation cloud 
(see Cowie \& McKee 1977).
It turns out that the cloud can survive from 1-13 Myr 
depending on the parameters of the wind.  

\par

\section{O{\sc vi} and X-ray Emission}

Figure 2 shows the O{\sc vi} and the X-ray maps of the model presented
here. As a gas element evaporates from the cloud, it quickly goes
through a large interval of temperatures, ranging from $T \sim 3\times
10^5$ K where the O{\sc vi} emission peaks, up to X-ray temperatures
(several $10^6$ K). Thus significant O{\sc vi} and X-ray emissions
originate close to the cloud surface, and are spatially
connected. For this reason their time evolution is connected to the
cloud dynamics, in particular to the variation of the cloud density at
$R_{\rm c}$ and to the evolution of the cloud size. 

\begin{figure}[ht]
\includegraphics[width=\textwidth]{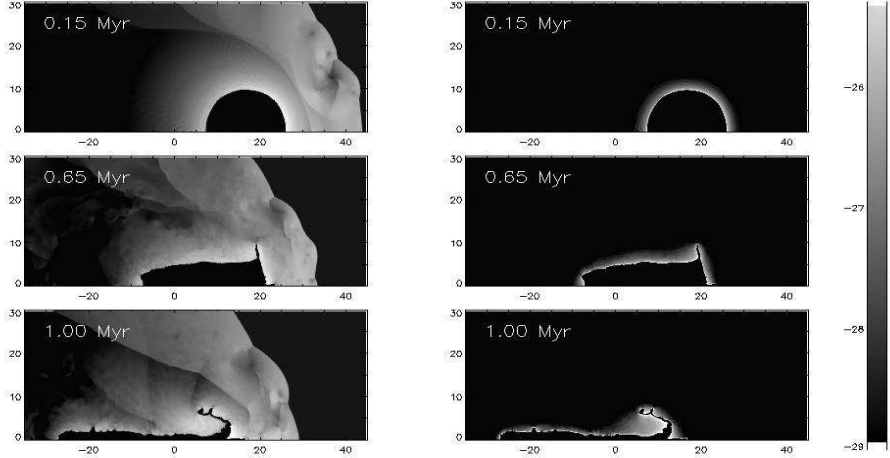}
\vskip.1in
\caption{X-ray (left panels) and O{\sc vi} (right panels) volume emissivity
at 3 different times. The logarithm of the soft X-ray (0.3 -- 2.0 keV
energy band) and O{\sc vi} volume emissivity (in units of erg s$^{-1}$
cm$^{-3}$) is shown.}
\end{figure}

The O{\sc vi} luminosity $L_{\rm OVI}$ oscillates in time between
$6\times 10^{32}$ and $10^{34}$ erg s$^{-1}$ with a period of 0.7 Myr
(see Fig. 3).  The X-ray luminosity $L_{\rm X}$ oscillates with the
same period, but to a lesser extent between $1.5\times 10^{33}$ and
$6\times 10^{33}$ erg s$^{-1}$. This lower amplitude is due to the
fact that the soft X-ray flux originates not only close to the cloud
surface, but also behind the bow shock; this second component is quite
steady and reduces the overall X-ray variability. We point out that
the O{\sc vi} and X-ray luminosities have a maximum at the leading
edge of the cloud close to the symmetry axis, where the density of the
evaporating gas is maximum. This is due in part to the larger
temperature of the shocked superwind, and partially to the compression
given by the ram pressure.
From Fig. 3 is clear that in the model in
which the effect of heat conduction is negletted, both $L_{\rm X}$ and
$L_{\rm OVI}$ are incresing with time according to the fragmentation
of the cloud.  \par 

In general only a small fraction ($0.1-0.4\%$) of the
mechanical energy of the wind that collides with the cloud is radiated
away from the cloud itself at FUV and X-ray wavelenghths. \par

Being the physical size of the observational $FUSE$ 30" aperture
dependent on the distance to the observed galaxies (equivalent to $\sim$
530 pc for M82, and for $\sim$ 2.5 kpc for NGC 3079), and that we do
not know how many clouds actually lie within these regions, our
quantitative comparison to the observational data will concentrate on
the ratio of O{\sc vi} to soft X-ray emission, rather than the
absolute emitted luminosities. Actually in our model with conduction
the ratio $L_{OVI}/L_{\rm X}$ assumes the values $\le 1.0$ that are
observed in NGC 4631 (Otte et al. 2003) and M82 (Hoopes et al. 2003)
for a considerable fraction of the time simulation. We stress that the
analogous model without thermal conduction gives a ratio
$L_{OVI}/L_{\rm X}$ much higher, largely out of the observed range.

\begin{figure}[ht]
\includegraphics[width=\textwidth]{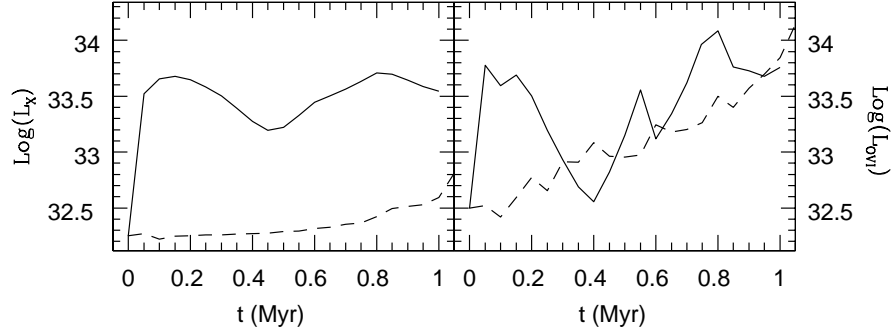}
\vskip.1in
\caption{Time evolution of the logarithm of X-ray (left panel) and 
O{\sc vi} (right panel) luminosities for the model with (solid line) 
and without (dashed line) heat conduction.}
\end{figure}

In addition to calculating the X-ray and O{\sc vi} emission we also
condsider absorption line properties, specifically column densities of
the O{\sc vi} ions probed in $FUSE$ observations of starbursts (see e.g.
Heckman et al. 2001, 2002). The mean value of log$(N_{OVI})$ through
the center of the cloud averaged over a radius of $R_{\rm sor}=5$ pc
is $\sim 13.19$ for the model without heat conduction and $\sim 13.43$
for the model with heat conduction. 

\section{Chemical Abundance of the X-ray Gas}

It is known that the abundance of the X-ray emitting gas in starburst
galaxies is rather puzzling. Although the superwind gas is supposed to
be composed mainly by SN ejecta, its metal abundance derived from its
X-ray emission is consistent with Solar abundance of the
$\alpha$-elements (Strickland et al. 2004a).  In our model the
emitting gas is composed by a mixture of evaporating cloud gas with
metallicity $Z_{\rm c}$ and enriched superwind material with
metallicity $Z_{\rm w}$. Thus the observed metallicity is given by
$Z_{\rm X}=Z_{\rm c}+A(Z_{\rm w}-Z_{\rm c}$), where $A=n_{\rm
w}/(n_{\rm c}+n_{\rm w}$) is weighted by the X-ray emission. $Z_{\rm
X}$ may be quite low for low values of $A$; this is actually the case
for this model where $A$ varies within the range 0.1-0.4 (i.e. the
observed abundance will be $Z_{\rm X} \sim Z_{\rm c}$, which should be
$\sim$ Solar). Much lower values of $A$ may be obtained varying some
parameters: for instance, for $v_{\rm w}=2236$ km s$^{-1}$ $A$ can be
as low as 0.02.

\end{document}